\documentclass[pre,twocolumn,a4paper,showpacs]{revtex4}
\usepackage{amsmath}
\usepackage{amssymb}

\usepackage{graphicx}

\usepackage{color} 

\makeatletter
\renewcommand{\@biblabel}[1]{\quad#1.}
\makeatother

\usepackage[tight,footnotesize]{subfigure}

\begin{document}

\title{Imitative learning as a connector of  collective brains}

\author{Jos\'e F. Fontanari}
\affiliation{Instituto de F\'{\i}sica de S\~ao Carlos,
  Universidade de S\~ao Paulo,
  Caixa Postal 369, 13560-970 S\~ao Carlos, S\~ao Paulo, Brazil}


\begin{abstract}
The notion that cooperation can aid a group of  agents to solve problems more efficiently than if those
agents worked in isolation  is prevalent, despite the little quantitative groundwork to support it. Here we consider a
primordial form of cooperation -- imitative learning -- that allows an effective exchange of information between agents, which are
viewed  as the processing units of a social intelligence system or collective brain.
In particular,
we use agent-based simulations to study the performance of a group of agents in solving a cryptarithmetic problem.
An  agent can either perform local random moves to explore the solution space of the problem or imitate a model agent -- the best performing
agent in its influence network.   
There is  a complex trade-off between the number of agents $N$ and the imitation probability $p$, and for the
optimal balance between these parameters we observe a thirtyfold diminution in the computational cost to find the  
solution of the cryptarithmetic problem as compared with the independent search. 
If those parameters are chosen far from the optimal setting, however, then  imitative learning can impair greatly the performance of the group.
The observed maladaptation of imitative learning for large $N$ offers an alternative  explanation for the group size of social animals.
\end{abstract}

\maketitle

\section{Introduction}

Imitative learning or, more generally, social learning offers a means whereby information can be transferred 
between biological or artificial agents, being thus a crucial factor for the emergence of social intelligence or collective brains \cite{Nehaniv_07}. 
Its relevance in this context is neatly expressed   by Bloom: ``Imitative learning acts like a synapse, allowing information to leap the gap from one creature to another''\cite{Bloom_01}. Not surprisingly, the advantages of this learning strategy  were perceived and exploited by nature well before  the advent of the human species as attested by its widespread use in the animal kingdom \cite{Moore_96,Fiorito_92,Laland_11,Heyes_12}. We note that imitation as a mechanism of social learning in humans was extensively studied by Bandura in the 1960s \cite{Bandura_62,Bandura_77}.

Social learning has inspired the  design of several optimization techniques, such as the 
particle swarm optimization algorithm \cite{Kennedy_95,Bonabeau_99} and the adaptive culture heuristic \cite{Kennedy_98,Fontanari_10}.
Despite the success of these heuristics in producing optimal or near optimal solutions to combinatorial optimization
problems, we know little about  the factors  that make  cooperation effective, as well as about the quantitative improvements  
that results from it \cite{Clearwater_91}. The reason is
probably that those  heuristics and the problems they are set to solve are too complex to yield to  a first-principle analysis.
In this contribution we  address these issues by  tackling a simple  combinatorial problem 
and by endowing the agents with straightforward search strategies in which the strength of collaboration is controlled by
a single parameter of the model. 
In particular, we solve a cryptarithmetic problem using a group of $N$ agents which, in addition to the  
capacity to carry out random  local searches,  can learn from
(or imitate) a model agent -- the best performing agent in their influence networks at a given trial.  The frequency of  the
imitative or cooperative behavior is determined by the imitation probability parameter  $p \geq 0$.  Hence our model exhibits  two critical ingredients of a
collective brain, namely,  imitative learning and a dynamic hierarchy among the agents  \cite{Bloom_01}.

We find  that imitative learning can greatly improve the performance of the group of agents provided that the 
control parameters are not too distant from their optimal values. For instance, in an optimal setting, say a fully
connected system of  $N=7$ agents with  imitation probability $p=0.6$, we find a  thirtyfold decrease
of the mean number of trials necessary to find the solution of the cryptarithmetic problem as compared with
the baseline case $p=0$ where the $N$ agents explore the solution space independently. Most significantly, 
however,  we find that, for a fixed value of the imitation probability, increasing   the number of agents $N$ beyond a
certain value impairs the group performance which can then perform much worse than in the case of 
the independent search. The following of a bad  model is the culprit for the poor performance in this case.
This harmful effect can be   mitigated somewhat by reducing  the connectivity of the agents so as to limit the influence of a model agent
to  only a fraction of the group. We argue that
the maladaptation of imitative learning for large systems  may be an alternative  explanation for the group size of social animals.

\section{Methods}

First we will describe   the particular cryptarithmetic problem
the agents must solve, explain how the digit-to-letter identifications are encoded in strings and introduce the cost value associated to those 
strings. We will present also the  elementary move that transforms any valid string
into an adjacent  valid string, allowing  thus  the exploration of the solution space. Once these basic elements are
introduced we will describe the mechanism of imitation between agents, thus completing the specification of the
agent-based model  we use to evaluate the efficacy of imitative learning in solving a complex task.

\subsection{The cryptarithmetic problem}

Cryptarithmetic problems such as 
\begin{equation}\label{DGR}
DONALD + GERALD = ROBERT 
\end{equation}
are constraint satisfaction problems in which the task is to find unique digit assignments to each of the letters so that the numbers represented by the words add up correctly \cite{Averbach_80}. In the cryptarithmetic problem of eq.\ (\ref{DGR}), there are $10!$ different digit-to-letter assignments, of which only one is the solution to the problem, namely, $A=4, B=3, D=5, E=9, G=1, L=8, N=6, O=2, R=7, T=0$.  This type of cryptarithmetic problem, in which the letters form meaningful words, are also termed alphametics \cite{Hunter_76} and were popularized in the 1930s by the {\it Sphinx}, a Belgian journal of recreational mathematics \cite{Averbach_80}. Of course, from the perspective of evaluating the performance of search
heuristics on solving cryptarithmetic problems, the meaningfulness of the words is inconsequential, but in this contribution we will focus mainly on the
alphametic problem (\ref{DGR}). Nonetheless we will offer evidence to support  the validity of our  conclusions by considering a few
randomly generated cryptarithmetic problems  as well.

A non-random search heuristics  to solve  cryptarithmetic problems requires  the introduction of some arbitrary quality measure or cost value
to each possible digit-to-letter  assignment.
For the alphametic problem of eq.\  (\ref{DGR}) we encode a digit-to-letter assignment by the string 
${\mathbf i} = \left ( i_1,i_2, \ldots, i_{10} \right )$ where
$i_n = 0, \ldots, 9$ represent the 10 digits and the subscripts $n=1, \ldots, 10$ label the letters according to the convention 
\begin{eqnarray}\label{convention}
1 & \to &  A \nonumber \\
2 & \to & B \nonumber \\
3 &\to  & D \nonumber \\
4 &\to  & E \nonumber \\
5 & \to & G \nonumber \\
6 &\to & L \nonumber \\
7 &\to &N \nonumber \\
8 &\to &O \nonumber \\
9 &\to &R \nonumber \\
10 &\to & T .
\end{eqnarray}
%
 For example, the
string $\left ( 0,2,9,4,8,1,7,6,3,5 \right)$ corresponds the the digit-to-letter assignment $A=0, B=2, D=9, E=4, G=8, L=1, N=7, O=6, R=3, T=5$.
A somewhat natural way to  associate a cost to a string ${\mathbf i}$ is through the expression \cite{Reza_91}
\begin{equation}\label{cost}
C \left ( {\mathbf i} \right ) = \left | R - \left ( F + S \right ) \right |
\end{equation}
where $R$ is the result of the operation ($ROBERT$), $F$ is the first operand ($DONALD$) and $S$ is the second operand ($GERALD$). 
In our example we have $R = 362435$, $F= 967019$ and $S = 843019$ so that the cost associated to string 
$\left ( 0,2,9,4,8,1,7,6,3,5 \right)$  is $C = 1447603$. If the cost of a string is $C=0$ then the digit-to-letter assignment coded
by that string is the  solution of the cryptarithmetic problem. We must note that the cost value defined in eq. (\ref{cost})
applies to all strings except those for which $i_3 = 0$ corresponding to the assignment $D=0$, $i_5 = 0$ 
corresponding to the assignment $G=0$ and $i_9 = 0$ corresponding to the assignment $R=0$. In principle, those are  invalid strings because
they violate the rule of the cryptarithmetic puzzles that an integer  should not have the digit $0$ at its leftmost position. For
those strings we assign an arbitrary large cost value, namely, $C = 10^8$, so that they can be considered valid strings and hence 
part of the solution space.

In addition to the assignment of the  cost values   to  
all  $10!$ strings that code the possible digit-to-letter mappings for the alphametic problem (\ref{DGR}), we
introduce also  an elementary  move that connects two valid digit-to-letter mappings. 
We define the elementary move as follows.
Starting from a particular digit-to-letter mapping, say $\left ( 0,2,9,4,8,1,7,6,3,5 \right)$, we choose
two letter labels at random and then  interchange the digits assigned to them. For example, say  we pick letter labels  $1$ and $5$,
then the resulting mapping after the elementary move is $\left ( 8,2,9,4,0,1,7,6,3,5 \right)$.
Clearly, the repeated application of our elementary move  
is capable of producing all $10!$  strings starting from any arbitrary valid digit-to-letter mapping. 

\subsection{Imitative learning}

The system is composed of $N$ agents or strings which represent valid digit-to-letter assignments as described before.  
Each agent is connected unidirectionally to exactly $M=1, \ldots, N-1$ distinct, randomly chosen
 agents in the system.  We will refer to those agents as the `influencers' of  the target agent.  More specifically, for each agent we sample $M$ influencers from the $N-1$ remaining agents without replacement. 
The extreme case $M=N-1$ corresponds to the fully connected network. 
An agent  has a probability $ p \in \left [0,1 \right )$ of copying 
a digit-to-letter assignment from a model string in its group of influencers, and probability $1-p$ of
performing the elementary move.  We choose the model string as the lowest cost string among the $M$ influencers of the target agent. 
If the cost associated to the target string  is  lower than the cost of the model string then the copying process is aborted. 

  To illustrate the  copying process  let us  assume for the sake of concreteness that the target agent  is our already 
familiar example string $\left ( 0,2,9,4,8,1,7,6,3,5 \right)$, whose cost is  $C = 1447603$, and that the model string is 
$\left ( 5,3,9,4,8,1,6,2,7,0 \right)$ whose cost is $C=1050568$. In the copying process the  target agent  selects at random one of the distinct digit-to-letter 
assignments in the model string and assimilates it. In our example, the distinct assignments occur at the letter labels $n=1,2,7,8,9,10$.  Say 
that the letter label $n=1$, which corresponds to the assignment $A=5$ according to  our convention (\ref{convention}), is chosen. 
To assimilate this assignment  the target agent needs to reassign the digit $0$ to the letter label  which was previously assigned to digit $5$ 
so that the resulting string becomes 
$\left ( 5,2,9,4,8,1,7,6,3,0 \right)$, whose cost  is $C=1448608$.  As expected, a result of the  imitative learning process is the increase of 
the similarity between the target and the model strings. 
The case $p=0$ corresponds to the  baseline limit where the $N$ agents perform  independent searches. The specific copying procedure proposed here was inspired by
the mechanism used to model the influence of an external media \cite{Shibanai_01,Avella_10,Peres_11} in Axelrod's model of culture dissemination \cite{Axelrod_97}.  It is important to note that in the case the target string is identical to the model string, as well as in the case
the cost of the target string is lower than the cost of the model string, the opportunity of update  is wasted.  

We may interpret the  imitation (or copying) process of  
a model string as a blackboard cooperation system  where a central control exhibits hints (i.e., the lowest cost string) in a public space 
\cite{Englemore_88,Clearwater_91}, but here we prefer to use the interpretation of learning by imitation as it allows us to view the group of
agents as a neural  system where the agents play the role of neurons and  dynamical synapses connect  them to the best performing
agents in the neural network  \cite{Bloom_01}. Nevertheless, since the process of imitation results in an effective collaboration among agents, in the sense that there is an  exchange of information between them, we refer to this search strategy as collaborative search to contrast with the
independent search  which  occurs when  the copying process is turned off, i.e, the imitation probability $p$ is set to zero.

\subsection{Search dynamics}

We begin by generating the $N$  influence networks, i.e.,  a group of $M$ influencers  for each
 agent. These networks are kept fixed during the entire search.  In this initial stage, at trial number $t=0$, we also associate  a 
random digit-to-letter assignment (a valid string)  to  each agent and determine its corresponding model string by evaluating and comparing the
cost values of its $M$  influencers. 

A new trial begins with the choice of the update order of the $N$ agents, so that at the end of  the trial all $N$ agents are updated.
The agent to be updated -- the  target agent --  has the possibility to imitate its model string  or perform the elementary move
 with probabilities $p$ and  $1-p$, respectively.
After update, we must re-evaluate the model string status in all groups of influencers  to which the target agent
belongs. After  all $N$ agents are updated we increment the trial number $t$ by one unit and check whether any
string has cost zero, in which case the search is halted. The trial number at which the search ends or, alternatively,
the number of trial to success is denoted by $t^*$.

Except for the independent search ($p=0$), the update of the $N$ agents is not strictly  a  parallel process
since the model strings may change several times within a given trial. Nonetheless, since in a single trial  all agents are
updated, the  total number  of  agent updates at trial $t$  is with given by the product $N t$.

\section{Results}

The efficiency  of a search strategy is measured by the total number of  agent updates necessary to find the solution 
of the cryptarithmetic problem (i.e., $N t^*$)
and in the following we will refer to this measure as the computational cost of the search. 
Since we expect that the typical number of trials   to success  $t^*$ scales with the size of the  solution space (i.e., $10!$),  we will present the results in terms of the  rescaled variable  
$\tau = t^*/10!$. For the
purpose of comparison we will consider first the baseline case of independent search for which the agents can perform the elementary
move only ($p=0$) and then the general case of cooperative search   ($p>0$).

\subsection{Independent search}

\begin{figure}[!ht]
\begin{center}
\includegraphics[width=0.48\textwidth]{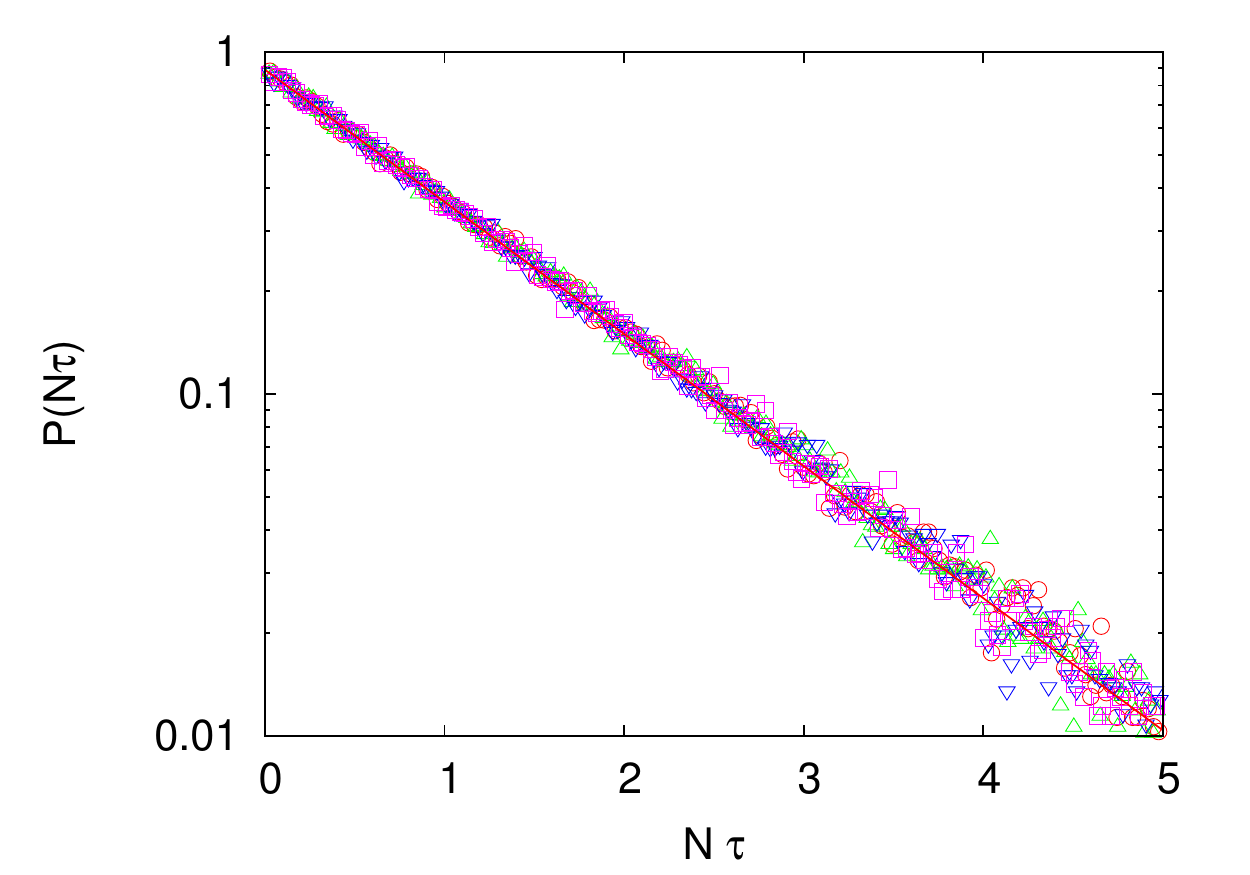}
\end{center}
\caption{
Exponential distribution of the rescaled computational cost for the independent search. Probability  distribution $P \left ( N \tau \right )$
that  a search employing $N$ independent agents  finds the  solution of the cryptarithmetic problem  (\ref{DGR}) using a total of  $N t^*$ updates
for  $N= 1 (\textcolor{red}{\circ}), 5 (\textcolor{green}{\vartriangle}), 10 (\textcolor{blue}{\triangledown})$ and $20 (\textcolor{magenta}{\Box})$. 
Here $\tau = t^*/10!$ is the ratio between number of trials to success and the size of the solution space.
These distributions were generated using $10^5$ independent searches for each $N$.
 The solid straight line is the exponential
distribution $P( N \tau) = a \exp \left ( - a N  \tau \right )$ with $ a =  1/1.14$. 
}
\label{fig:1}
\end{figure}

In this case there is no imitation and so the influence networks have no role in the outcome of the search. The  main results of the independent search are summarized in fig.\ \ref{fig:1}, which shows the probability distribution  $P \left (N  \tau \right) $ of  the rescaled computational cost  $N \tau$ of the search
for several system sizes. The data is very well fitted by the exponential distribution $ P \left ( N \tau \right ) =  a \exp \left ( - a N \tau \right )$
with $ a = 1/1.14$ which is shown by the solid straight line in the figure.

As expected, the mean rescaled computational cost $\langle N \tau  \rangle \approx 1.14$ is insensitive to the system size $N$, but the finding that it does not equal 1
is somewhat surprising. In fact, if we replace our elementary  move by a global move in which the entire string 
is generated randomly  at each update then we find that  this mean equals 1, as expected. The reason that our elementary move is
slightly less efficient than the global move in exploring the solution space  is because  it  is not too unlikely to reverse a change  made
by the elementary move. For example, the probability to reverse a change  in a subsequent trial is  $2/10 \times 1/9 = 2/90$ for the
 the elementary move, whereas it  is $1/10!$ for the global move.

\subsection{Cooperative search}

As pointed out before, the cooperation among agents stems from the possibility that they copy potentially relevant digit-to-letter assignments from
the model strings in their influence networks. We will consider first the fully connected system where $M=N-1$ and then the 
partially connected systems  where $1 \leq M < N-1$.

\subsubsection{Fully connected system}

\begin{figure}[!ht]
\begin{center}
\includegraphics[width=0.48\textwidth]{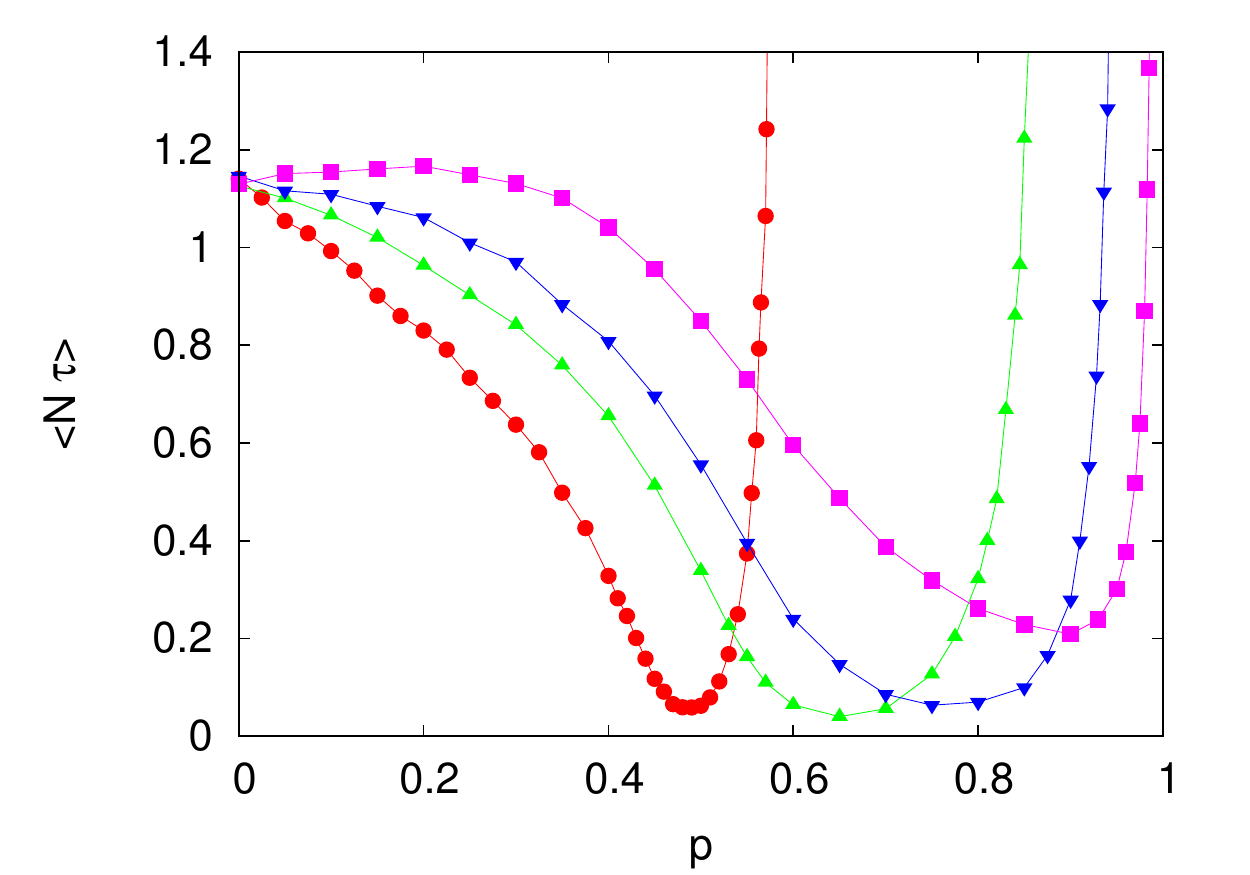}
\end{center}
\caption{
The effect of the imitation probability on the computational cost of the fully connected system. The symbols
represent the mean rescaled computational cost $ \langle N \tau \rangle$ for cooperative systems of size 
$N = 20 (\textcolor{red}{\bullet}),  N = 5 (\textcolor{green}{\blacktriangle}), N=3 (\textcolor{blue}{\blacktriangledown})$
and $N=2 (\textcolor{magenta}{\blacksquare})$. The independent variable $p$ is the probability that
an agent will copy a digit-to-letter assignment from the model string, chosen as the lowest cost string in
the entire system. Each symbol represents an average over $10^5$  searches and the lines
are guides to the eye. The error bars are smaller than the size of the symbols.
}
\label{fig:2}
\end{figure}

Figure \ref{fig:2} shows how  the mean rescaled computational cost is affected by varying the imitation probability $p$  while the number of agents $N$ is kept at a fixed value.
For $N=20$  and $p=0.5$  we observe a twentyfold decrease of the mean cost in
comparison with the cost of the independent search, which corresponds to $p=0$ and yields $  \langle N \tau \rangle \approx 1.14$.
This is a remarkable evidence of the power of imitative learning to speed up the search on the solution space of the cryptarithmetic problem.
In the limit $p \to 1$ one expects the computational cost to diverge since the solution space cannot be fully explored  
as the option for the elementary move is never made in this limit.  This harmful effect of learning by imitation becomes more
pronounced as the number of agents  increases.

\begin{figure}[!ht]
\begin{center}
\includegraphics[width=0.48\textwidth]{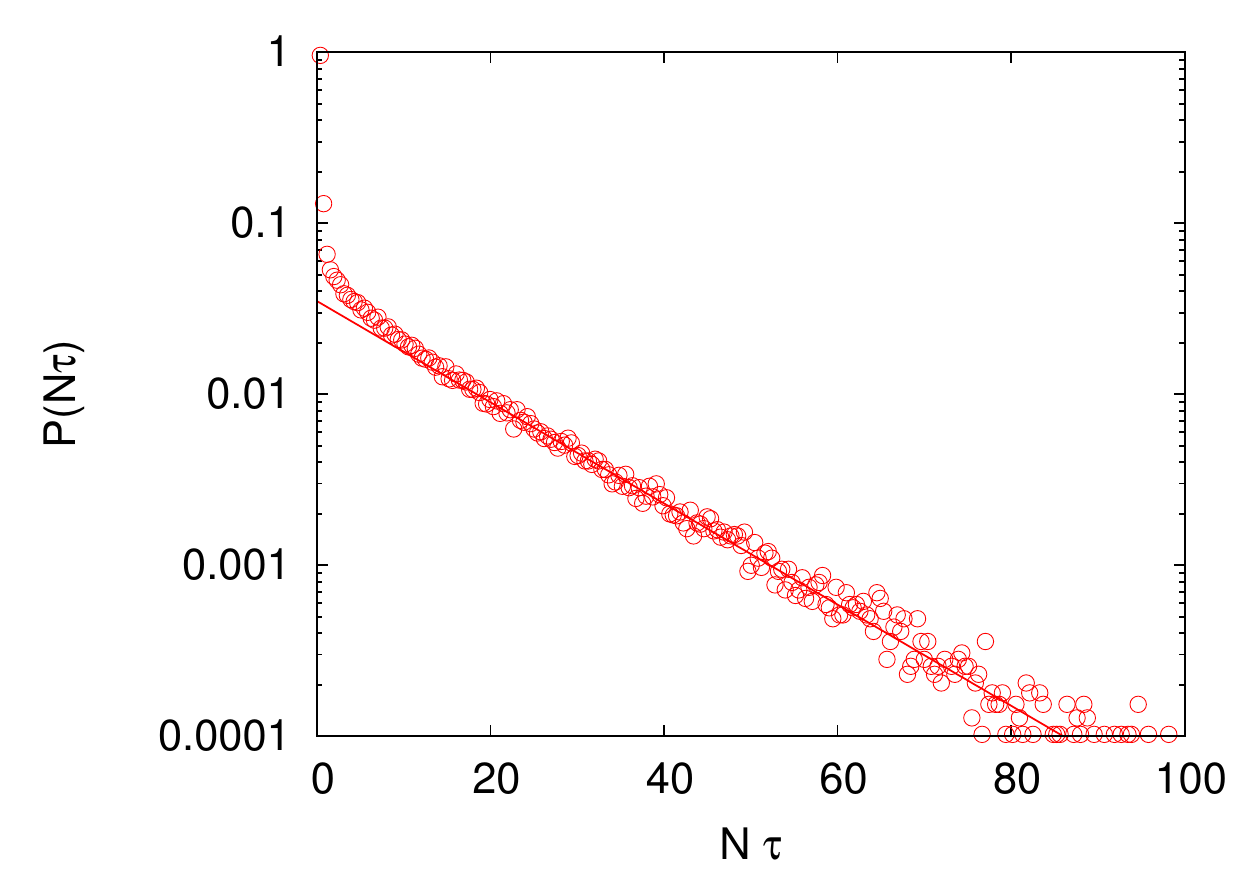}
\end{center}
\caption{
 Deviation from the exponential distribution for a large imitation probability.  
Probability  distribution $P \left ( N \tau \right ) $ of the rescaled computational cost
for a search employing $N=20$  fully connected agents  with imitation probability $p=0.6$. The
mean of this distribution is $\langle N \tau \rangle \approx 8.0$.
 The solid straight line is the fitting function $ a \exp \left ( -b  N  \tau \right )$ with $ a = 0.03$ and $b= 1/15$ in the
regime of large cost.  The distribution was generated using $10^5$ independent searches.
}
\label{fig:3}
\end{figure}

In the region where the mean computational cost decreases monotonically with increasing $p$ (e.g.,  $p < 0.5$ for $N=20$) 
we  found that the probability distribution of the computational cost  is well described by an exponential distribution, in the sense
that the ratio between the standard deviation and the mean is always very close to 1. (We recall that this ratio  equals 1 for an exponential distribution.)
However, in the region where $  \langle N \tau \rangle $ increases with increasing $p$ we found that  in the low cost regime $P \left ( N \tau \right )$ gives values  significantly greater than those predicted by an exponential distribution, as illustrated in fig.\ \ref{fig:3}, though those values
are not  greater than those obtained in the case of the independent search (see fig.\ \ref{fig:1}).

\begin{figure}[!ht]
\begin{center}
\includegraphics[width=0.48\textwidth]{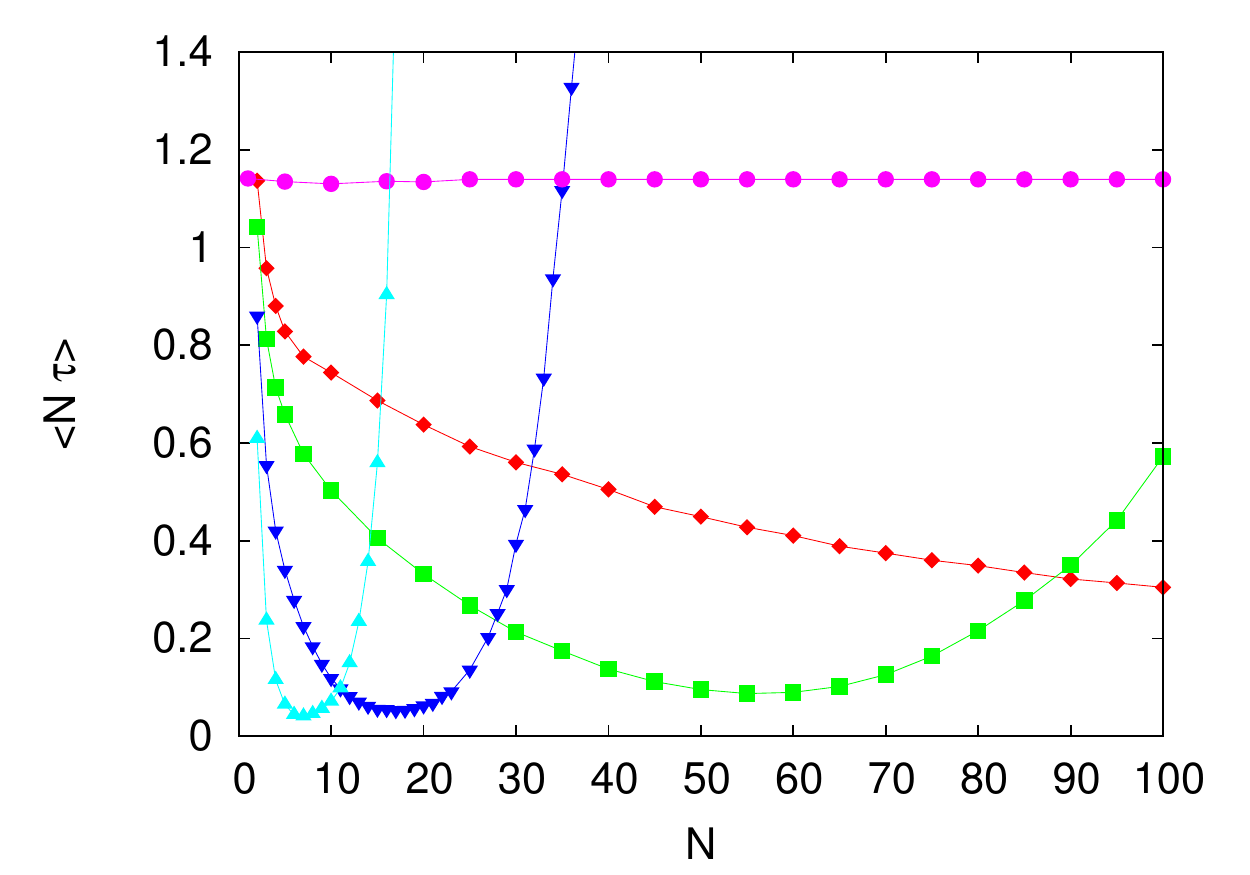}
\end{center}
\caption{
The effect of the system size on the computational cost of the fully connected system.  
The symbols
represent the mean rescaled computational cost $ \langle N \tau \rangle$ for the imitation probability 
$p = 0 (\textcolor{magenta}{\bullet}), p=0.3 (\textcolor{red}{\blacklozenge}),
p= 0.4 (\textcolor{green}{\blacksquare}), p=0.5 (\textcolor{blue}{\blacktriangledown})$ 
and $p = 0.6 (\textcolor{cyan}{\blacktriangle})$
 The independent variable $N$ is the number of agents in the system. Each symbol represents an average over $10^5$  searches and the lines are guides to the eye. The error bars are smaller than the size of the symbols.
}
\label{fig:4}
\end{figure}

The effect of increasing the number of agents $N$ for a fixed value of the imitation probability  $p$ is summarized in fig.\ \ref{fig:4}. 
The mean computational cost of the cooperative system exhibits a  non-monotonic dependence on $N$, except in the case of the independent search ($p=0$) 
when it takes on a constant value. The benefit of cooperation is  seen in this figure by the initial decrease of the 
computational cost
as the number of agents increases. However, for all $p>0$ we find that the presence of  too many agents    greatly harms the performance of the system and that for
a fixed $p>0$ there exists an optimum value of $N$ that maximizes the search efficiency of the cooperative system. For instance, although not shown in
the scale of  fig.\ \ref{fig:4}, the
minimum computational cost for $p=0.3$ occurs at $N \approx 270$. The efficiency at this
optimum, however, is not  affected significantly by the choice of the parameters $N$ and $p$. In other words, the costs corresponding to the minima shown
in figs.\ \ref{fig:2} and \ref{fig:4} are not very  sensitive to changes in $N$ and $p$, respectively.
 In particular, for the parameter settings we have 
explored, the best efficiency $ \langle N \tau \rangle \approx 0.041$ is achieved for $N=7$ and $p=0.6$ and amounts to a thirtyfold speed up
with respect to the independent search.

We conjecture that the reason the efficiency of the cooperative system deteriorates as $N$ increases beyond its
optimum value (e.g., in the range $N > 6$ for $p=0.6$ as shown in fig.\ \ref{fig:4}) is that for $N$ not too small there is a good 
chance that the cost of one of the  strings is significantly lower than the cost of the other $N-1$ strings. Provided $p$
is not too  small too, this string may remain as the model string  for a few trials thus
biasing the search to the vicinity of  the model string. In the (typical)  case that the model string is  far from the
solution of the cryptarithmetic problem,  imitative learning may lead to the observed impairment of the performance of the cooperative system. In sum, 
the following of
a bad leader is likely the culprit of the poor performance of the system.  

\begin{figure}[!ht]
\begin{center}
\subfigure{\includegraphics[width=0.48\textwidth]{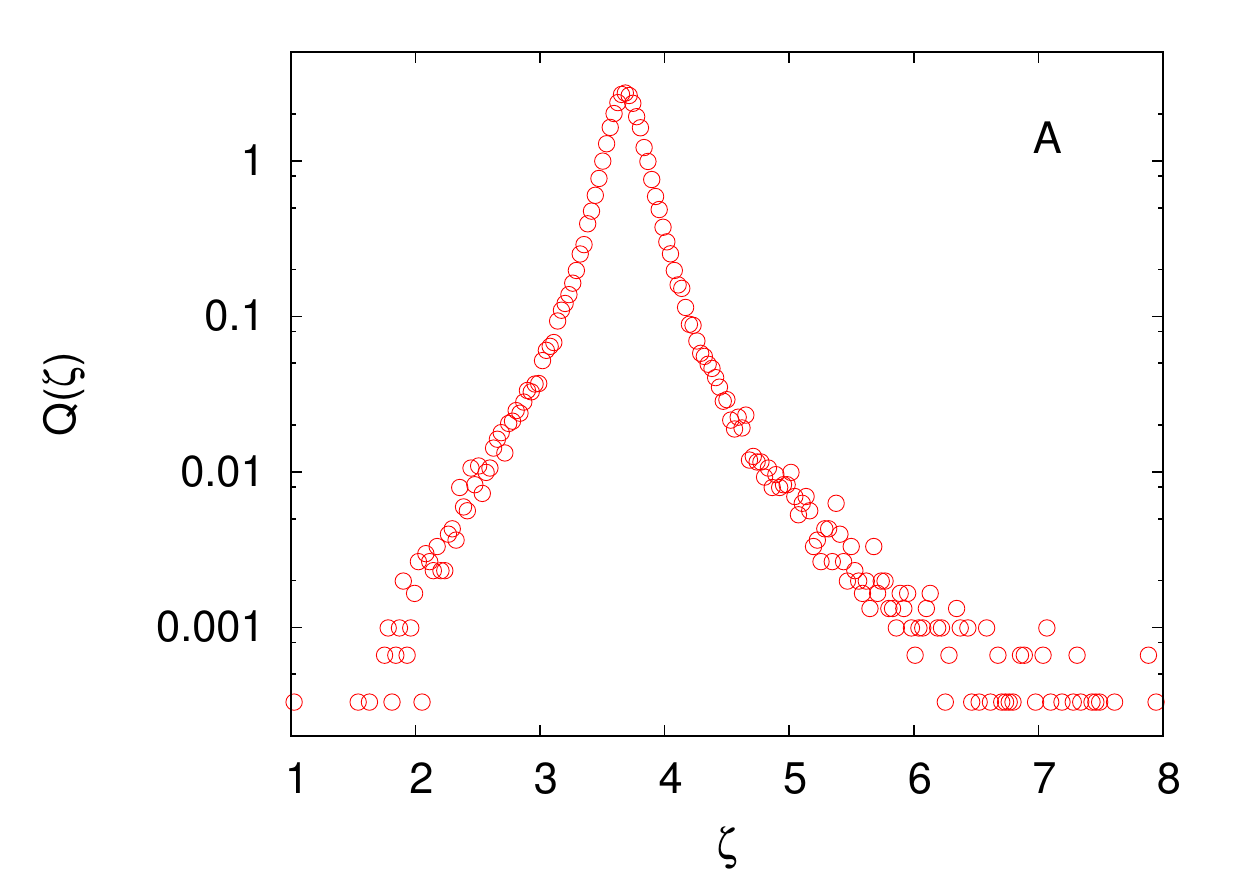}}
\subfigure{\includegraphics[width=0.48\textwidth]{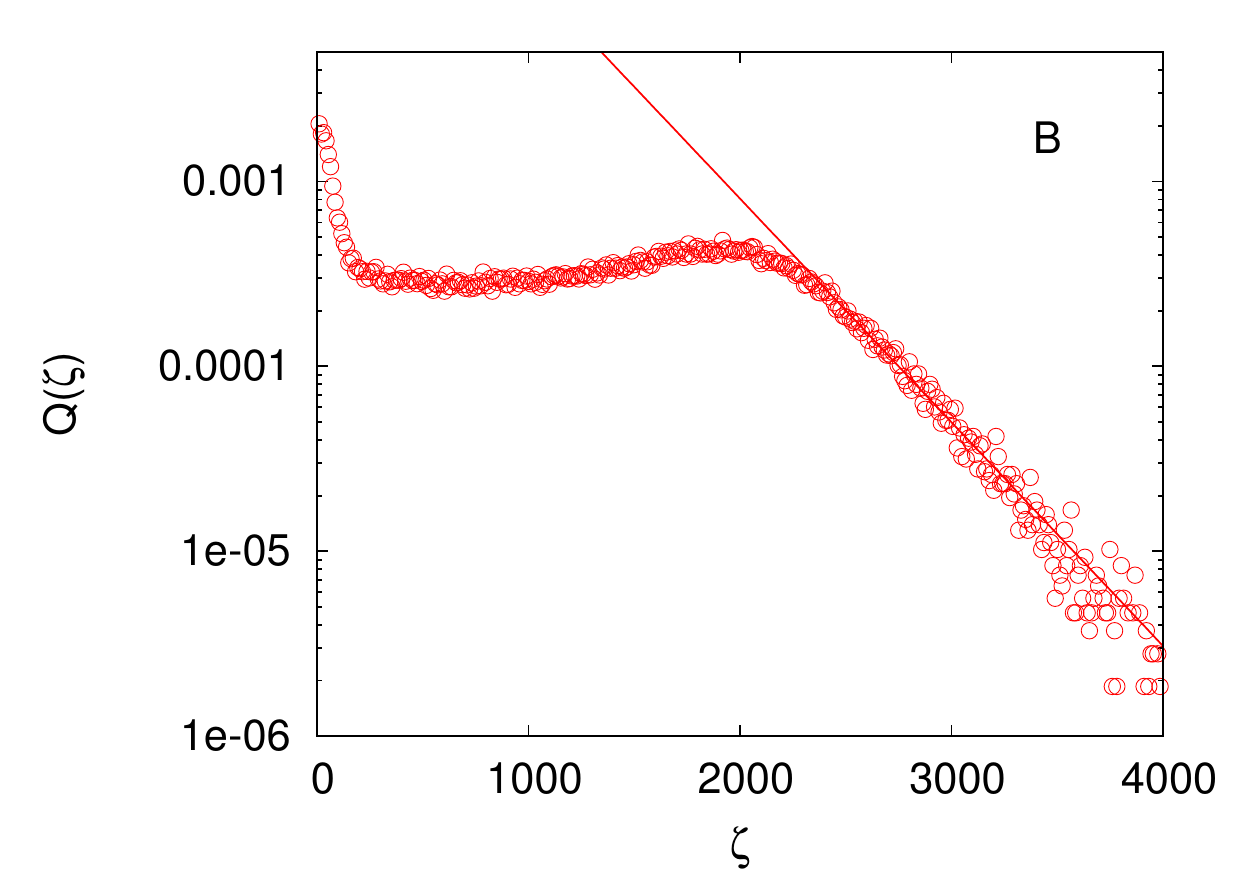}} 
\end{center}
\caption{
Probability distribution  of the mean duration of the stases in a search.
Probability distribution  of the mean number of trials  $\zeta$ for which a cost value
stays as the lowest cost  among the $N$  solutions in  $10^5$ searches
for the imitation probability  $p=0.6$  in a fully connected system. Panel  {\bf A}:  $N= 6$ (low computational cost regime). 
Panel {\bf B}:  $N=17$ (high computational cost regime). The slope of the straight line shown  in the semi-log scale of panel {\bf B}  is $0.003$.
}
\label{fig:5}
\end{figure}

To check the validity of this conjecture  we calculate  the mean number of consecutive trials for which a cost value
stays as the lowest cost  among the $N$  strings. The procedure to obtain this quantity, which we denote by $\zeta$,  is 
straightforward. At trial $t=0$ we evaluate the cost of the $N$ strings and record the minimal cost among them.
Then at the next trial $t=1$, after the $N$ strings
are updated, we re-evaluate again their costs and record the minimal cost. If the minimal cost at $t=1$ is different, i.e., greater or less,  than the
minimal cost at $t=0$ we  say that a change  event has occurred. The comparison of the values of the minimal costs at consecutive
trials  is repeated  and the cumulative number of change events is recorded until the  solution is found at $t=t^*$. 
The desired quantity  $\zeta$ is given simply by the ratio between the total number of change events and the total number  of
trials  $t^*$. Hence for each search we obtain a single value for $\zeta$, which  can then be interpreted as the mean number of trials between
consecutive  change events  or as the mean duration of the stases for that search.

In fig.\ \ref{fig:5} we present
the probability distribution $Q \left ( \zeta \right )$ using  $10^5$  searches for the imitation probability $p=0.6$ and
two representative values of $N$.   Figure \ref{fig:5}{\bf A} shows this distribution for $N=6$, which corresponds to a regime of low computational
cost  according to fig.\ \ref{fig:4}. We observe a pronounced maximum at $\zeta \approx 3.7$  so that
in most searches  the model cost remains unaltered for 3 to 5 trials. This is an optimum scenario since no  string  stays
on the top tier long enough to influence the entire system. For $N < 6$, we find that $Q \left ( \zeta \right )$ exhibits a similar shape 
but  the maximum becomes sharper and its location is shifted towards lower values of $\zeta$ as 
$N$ decreases.
Figure  \ref{fig:5}{\bf B}, which  shows the results for $N=17$,  reveals a very different scenario: the distribution $Q \left ( \zeta \right )$ exhibits 
a plateau  indicating that
the model cost remains unchanged for hundreds to a few thousands trials. For very large values of $\zeta$, the distribution $Q \left ( \zeta \right )$ 
seems to exhibit an  exponential decay  to zero, namely, $Q \left ( \zeta \right ) \sim \exp \left ( -0.003 \zeta \right )$.   We stress that for the two cases  exhibited in fig.\ \ref{fig:5} the probability that an agent will imitate the model rather than perform an elementary move is the same, namely $p=0.6$, and so the qualitative differences reported in the figure are due solely  to the  change on the number of agents.

\subsubsection{Partially connected system}

\begin{figure}[!ht]
\begin{center}
\includegraphics[width=0.48\textwidth]{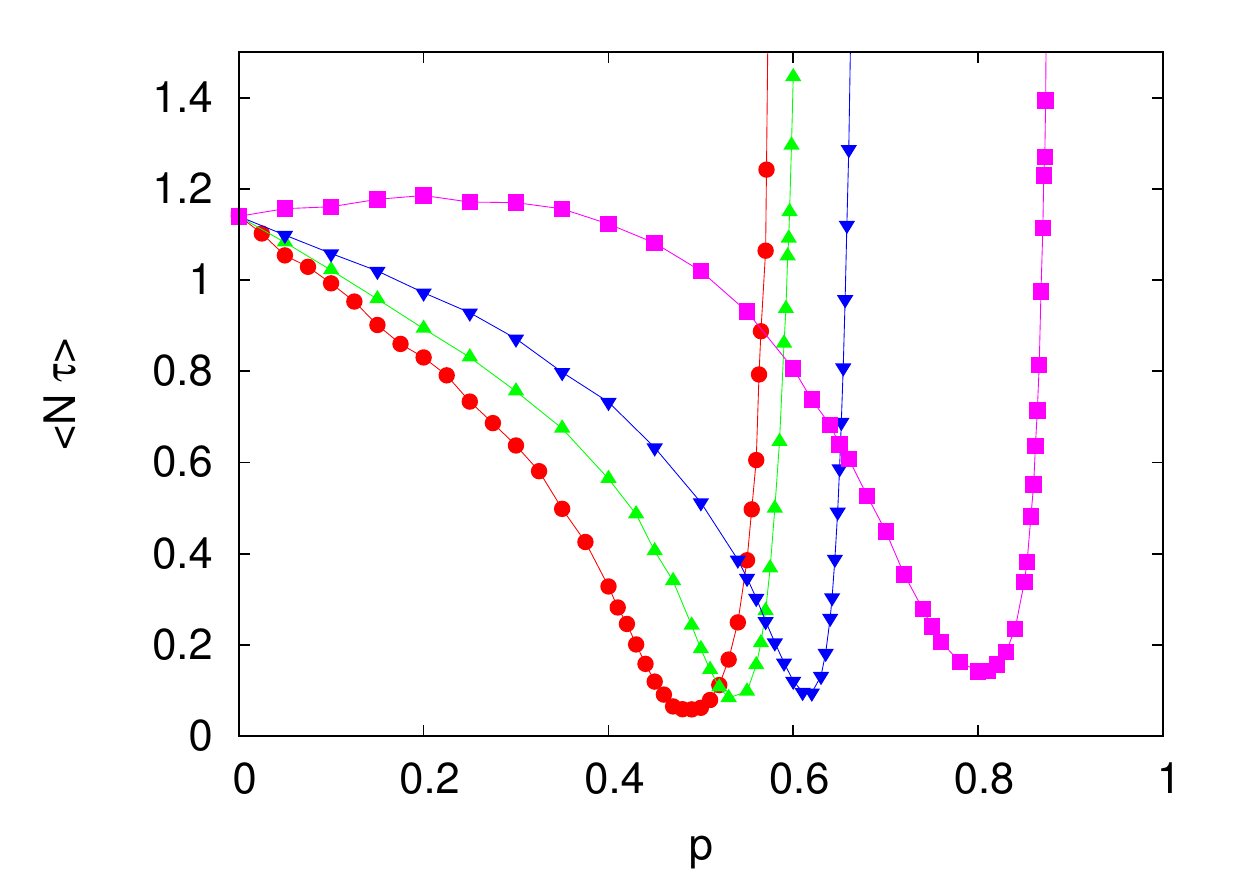}
\end{center}
\caption{
The effect of the imitation probability on the computational cost  of partially connected systems. The symbols
represent the mean rescaled computational cost $ \langle N \tau \rangle$ for a system
composed of $N=20$ agents, each one connected
to $M = 19 (\textcolor{red}{\bullet}),  M = 9 (\textcolor{green}{\blacktriangle}), M=4 (\textcolor{blue}{\blacktriangledown})$
and $M=1 (\textcolor{magenta}{\blacksquare})$ influencers. The independent variable $p$ is the imitation probability. 
Each symbol represents an average over $10^5$  searches and the lines
are guides to the eye. The error bars are smaller than the size of the symbols.
}
\label{fig:6}
\end{figure}

\begin{figure}[!ht]
\begin{center}
\includegraphics[width=0.48\textwidth]{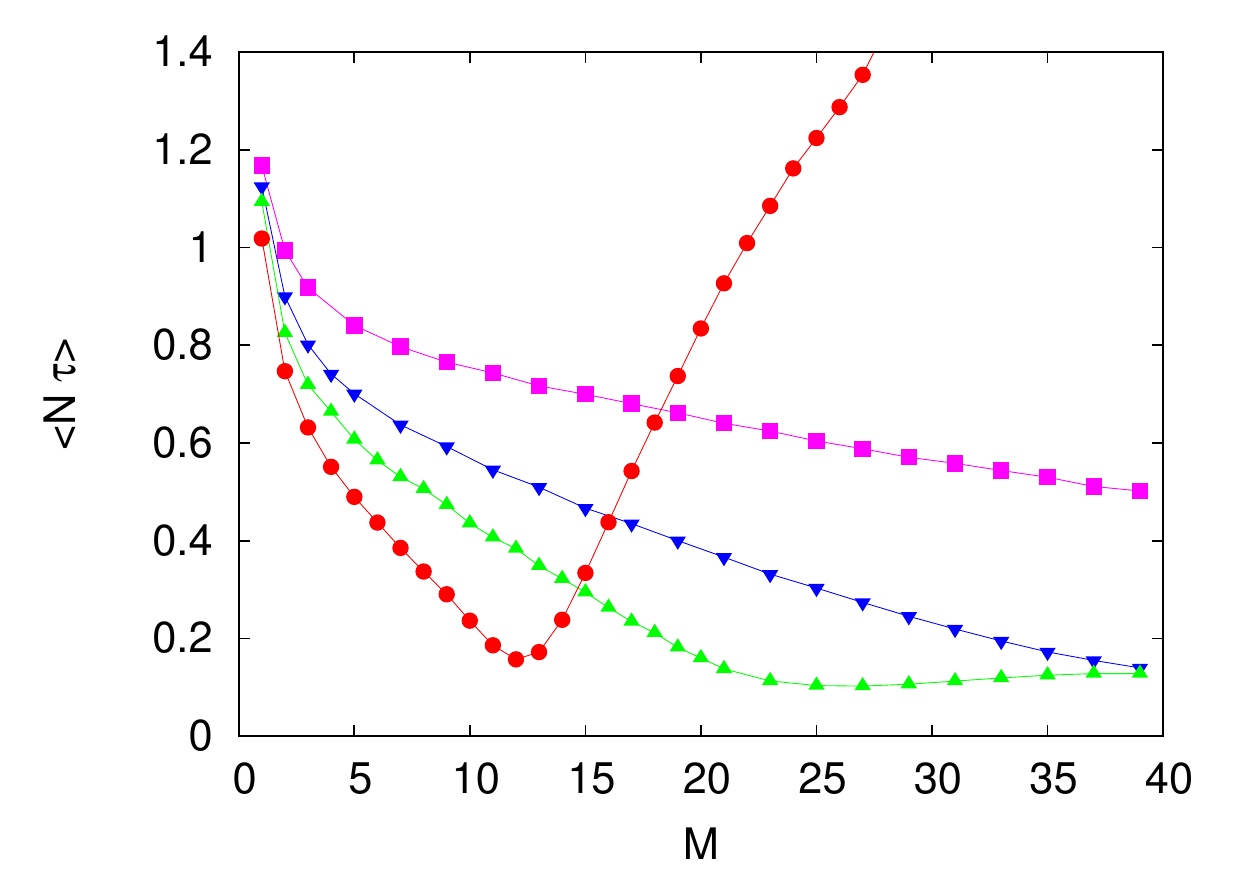}
\end{center}
\caption{
The effect  of the number of  influencers on the mean computational cost.  The symbols
represent the mean rescaled computational cost $ \langle N \tau \rangle$ for a system
composed of $N=40$  agents and imitation probability
$p = 0.5 (\textcolor{red}{\bullet}),  p= 0.45 (\textcolor{green}{\blacktriangle}), p=0.4 (\textcolor{blue}{\blacktriangledown})$
and $p=0.3 (\textcolor{magenta}{\blacksquare})$. The independent variable $M$ is the size of the group of
influencers of each agent. Each symbol represents an average over $10^5$  searches and the lines
are guides to the eye. The error bars are smaller than the size of the symbols.
}
\label{fig:7}
\end{figure}

If the poor performance of large collaborative systems based on imitative learning is due to  the influence of bad models then a natural
way to  reduce this harmful effect  is to limit the influence of those models. This was the motivation to introduce the influence networks scheme where 
each  agent picks its model among $M$ randomly chosen agents predetermined  at the beginning of the search. In fact, fig.\ \ref{fig:6} shows that the reduction
of the connectivity of the  agents increases somewhat the range of values of the imitation probability $p$ for which the cooperative system outperforms the system composed of independent agents.
More pointedly, for $N=20$ this range is extended from $p \approx 0.57$ for
$M=19$ to $p \approx 0.87$ for $M=1$. In addition, the value of the optimal  mean computational cost does not seem to vary significantly with $M$.
Figure  \ref{fig:7} offers  another perspective on the role of the number of influencers $M$. It  shows that
for small values of the imitation probability the fully connected system (i.e., $M=N-1$) exhibits the best performance. However, 
as $p$ increases (e.g., $p> 0.4$ for $N=40$), the optimal performance is obtained with partially connected systems. Moreover,
we found that for any fixed  value of $p>0$ and $M$ the performance of the system is always impaired  when the number of agents $N$
is very large.
Finally, we note that similarly to our findings for the fully connected system, the probability distribution of the computational cost $P \left ( N \tau \right )$ departs significantly
from an exponential distribution only in the regions where the mean computational cost becomes an increasing function of the
control parameters of the model.

\subsubsection{Random cryptarithmetic problems}

\begin{figure}[!ht]
\begin{center}
\includegraphics[width=0.48\textwidth]{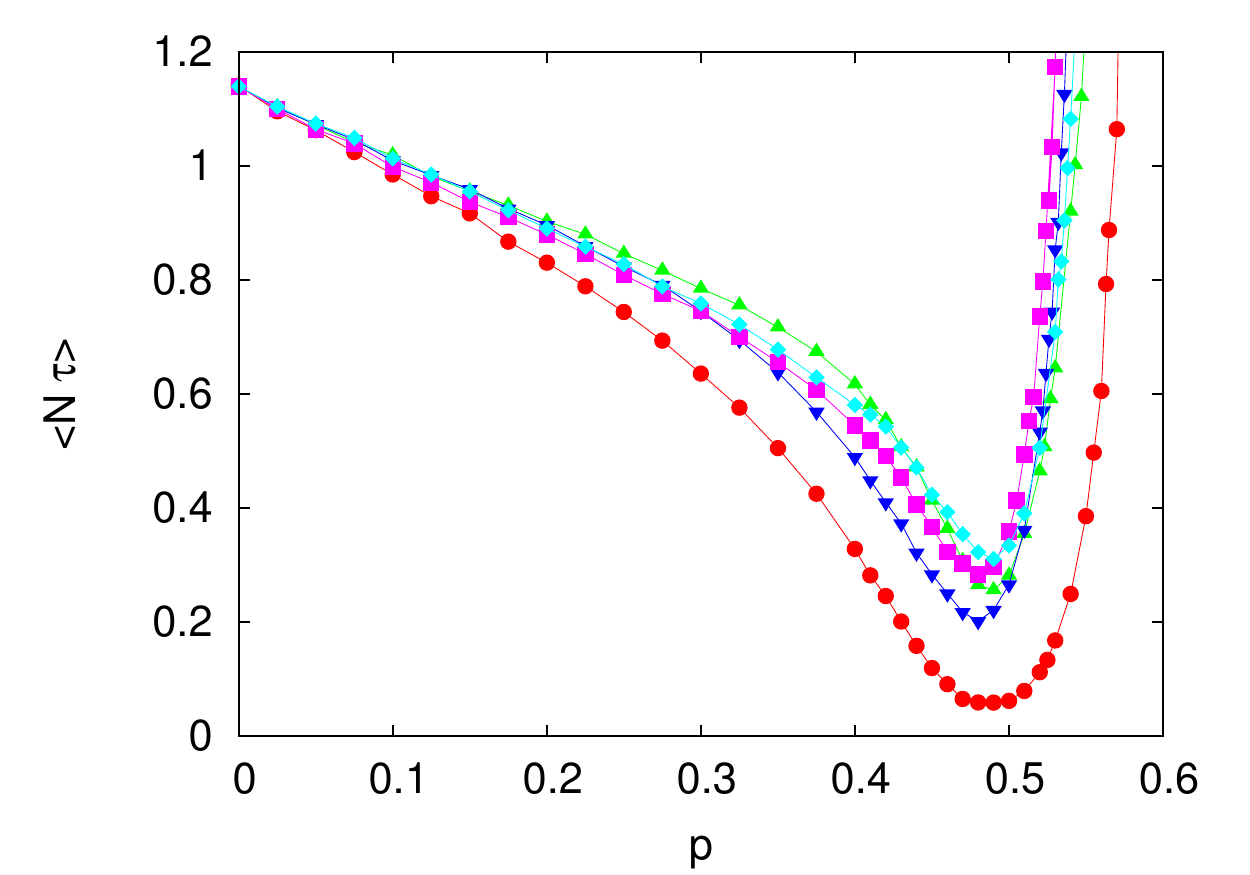}
\end{center}
\caption{
Computational cost of the alphametic problem and of four random
cryptarithmetic problems. 
The mean rescaled computational cost for the alphametic problem $DONALD + GERALD = ROBERT$ (symbol $\textcolor{red}{\bullet}$)
and for four ten-letter random cryptarithmetic problems  with a unique solution (symbols $\textcolor{blue}{\blacktriangledown},
\textcolor{green}{\blacktriangle}, \textcolor{magenta}{\blacksquare}$ and  $\textcolor{cyan}{\blacklozenge}$).
The symbols
represent the mean rescaled computational cost $ \langle N \tau \rangle$ for a system
composed of $N=20$  fully connected agents. 
 The independent variable $p$ is the imitation probability. Each symbol represents an average over $10^5$  searches and the lines
are guides to the eye. The error bars are smaller than the size of the symbols.
}
\label{fig:8}
\end{figure}

In order to verify the generality of our findings, which were obtained for the specific  alphametic problem  $DONALD + GERALD = ROBERT$,
we have considered a variety of random cryptarithmetic problems with 10 letters  and  a unique solution, so that the sizes of their solution spaces
are the same as that of the alphametic problem. The comparison between the mean computational costs to solve 
four such random problems and our alphametic  problem  is shown in fig.\ \ref{fig:8} for the fully connected system. The results are qualitatively the same, as expected. The alphametic problem, however,
was somewhat easier to solve by the cooperative system than the random problems, perhaps because of the coincidence of the
last three letters (``ALD")  in the first and second operands. Interestingly, the independent system ($p=0$) cannot distinguish between the problems but
the cooperative system ($p>0$) can, and this distinction is most pronounced when the parameters are set so as to achieve the optimal performance.
It is as if the cooperative system had adapted to the specific task posed to it. We expect that our conclusions 
remain  valid, in a qualitative sense of course, for any  constraint satisfaction problem characterized by a  rugged cost landscape.  

\section{Discussion}

Rather than offer any  novel  method to solve cryptarithmetic problems,  our aim in this contribution is  to assess quantitatively the potential
of imitative learning as the underlying mechanism   -- the critical connector --  of collective brains \cite{Bloom_01}. Here imitative learning is implemented by allowing an agent to copy clues  from the best performing agent -- the model agent -- in its group of influencers.
More pointedly, at  trial $t$ each agent has the probability $p$ of  imitating the model and the probability $1-p$ of executing a random rearrangement 
of the digit-to-letter mapping  which is its guess to the solution of the cryptarithmetic problem.
In an optimal setting, say  a fully connected system of $N=7$ agents with imitation probability $p=0.6$, we find a  thirtyfold decrease of the 
mean number of trials needed  to find the solution of the problem (i.e., of the mean computational cost), as compared with the case $p=0$ when  the agents search  the solution space independently (see fig.\ \ref{fig:4}). 

In the optimal setting, as well as in the regions where the computational cost is a decreasing function of the control parameters of the model,
the probability distribution of the computational cost
is given by an exponential distribution, rather than by  a lognormal  distribution  as predicted by a general theory of cooperative
processes \cite{Clearwater_91,Huberman_90}. In fact, the reason the cooperative scheme implemented in  \cite{Clearwater_91} 
is so efficient is that all discovered digit-to-letter assignments that add up correctly modulo 10 for at least one column are permanently
exposed as hints in a blackboard  for use by all agents, which can pick a hint at each trial. 
There is no place for any kind of learning in that scenario since  in the case there are no hints in the blackboard or  the agent 
has already used the chosen one, the agent selects a new random digit-to-letter assignment which is completely uncorrelated to its previous 
assignment. 

Most significantly, for  fixed values of the imitation probability $p$ 
and of the number of influencers $M$, we find that increasing the number of agents $N$ beyond a certain quantity 
impairs the working of the cooperative system, which then  performs much worse than if the agents  had executed independent searches. 
Our analysis indicates that  the following of  a bad model is  the culprit of the poor performance of the system in this case. 
In that sense, the efficacy of imitative learning could be a  factor determinant of group size \cite{Wilson_75}. In contrast to
the cognitive load that constrains the number of individuals with whom it is possible to maintain stable relationships
and leads to Dunbar's number for primates \cite{Dunbar_92,Vespignani_11}, the group size here (i.e., the value of $N$ corresponding
to the lowest computational cost) does not stem from a limitation of the neocortical processing capacity of the individuals. Rather, it is a
property of the group of agents as a whole, since for any fixed non-vanishing value of the imitation probability, which may be seen as an individual 
trait,  a too large number of agents, which is a group property,   will impair the performance of the cooperative system.  Of course, if  $p$ were allowed
to decrease with increasing $N$ then the system could be maintained at the highest level of perform regardless of the group size
(see figs.\ \ref{fig:2} and \ref{fig:4}). In other words, in order to perform at the optimal level a system based on imitative learning should
decrease the frequency of  the interactions among individuals as its size increases.

To conclude, our findings indicate that imitative learning has a great potential to improve the task-solving capability of
a group of agents, provided the model parameters -- number of agents ($N$),  imitation probability ($p$) and 
number of influencers ($M$) --  are not too far from their optimal values. 
In the cases that  $N$ or $p$ are too large,
the imitative learning strategy leads the  cooperative system astray, in a sort of maladaptive behavior that has actually been observed 
  in fishes \cite{Laland_98}. It would be interesting to find out what new ingredients
one should add to our  model  in order to prevent the  catastrophic effect of imitative learning  on large populations.

\acknowledgments

This work was partially supported  by  grant    2013/17131-0, S\~ao Paulo Research Foundation (FAPESP)
and by grant  303979/2013-5, Conselho Nacional de Desenvolvimento Cient\'{\i}fico e Tecnol\'ogico (CNPq).

\bibliography{arX_fontanari}

\begin{thebibliography}{10}%
\makeatletter
\providecommand \@ifxundefined [1]{%
 \ifx #1\undefined \expandafter \@firstoftwo
 \else \expandafter \@secondoftwo
\fi
}%
\providecommand \@ifnum [1]{%
 \ifnum #1\expandafter \@firstoftwo
 \else \expandafter \@secondoftwo
\fi
}%
\providecommand \enquote [1]{``#1''}%
\providecommand \bibnamefont  [1]{#1}%
\providecommand \bibfnamefont [1]{#1}%
\providecommand \citenamefont [1]{#1}%
\providecommand\href[0]{\@sanitize\@href}%
\providecommand\@href[1]{\endgroup\@@startlink{#1}\endgroup\@@href}%
\providecommand\@@href[1]{#1\@@endlink}%
\providecommand \@sanitize [0]{\begingroup\catcode`\&12\catcode`\#12\relax}%
\@ifxundefined \pdfoutput {\@firstoftwo}{%
 \@ifnum{\z@=\pdfoutput}{\@firstoftwo}{\@secondoftwo}%
}{%
 \providecommand\@@startlink[1]{\leavevmode\special{html:<a href="#1">}}%
 \providecommand\@@endlink[0]{\special{html:</a>}}%
}{%
 \providecommand\@@startlink[1]{%
  \leavevmode
  \pdfstartlink
   attr{/Border[0 0 1 ]/H/I/C[0 1 1]}%
   user{/Subtype/Link/A<</Type/Action/S/URI/URI(#1)>>}%
  \relax
 }%
 \providecommand\@@endlink[0]{\pdfendlink}%
}%
\providecommand \url  [0]{\begingroup\@sanitize \@url }%
\providecommand \@url [1]{\endgroup\@href {#1}{\urlprefix}}%
\providecommand \urlprefix [0]{URL }%
\providecommand \Eprint[0]{\href }%
\@ifxundefined \urlstyle {%
  \providecommand \doi [1]{doi:\discretionary{}{}{}#1}%
}{%
  \providecommand \doi [0]{doi:\discretionary{}{}{}\begingroup
  \urlstyle{rm}\Url }%
}%
\providecommand \doibase [0]{http://dx.doi.org/}%
\providecommand \Doi[1]{\href{\doibase#1}}%
\providecommand \bibAnnote [3]{%
  \BibitemShut{#1}%
  \begin{quotation}\noindent
    \textsc{Key:}\ #2\\\textsc{Annotation:}\ #3%
  \end{quotation}%
}%
\providecommand \bibAnnoteFile [2]{%
  \IfFileExists{#2}{\bibAnnote {#1} {#2} {\input{#2}}}{}%
}%
\providecommand \typeout [0]{\immediate \write \m@ne }%
\providecommand \selectlanguage [0]{\@gobble}%
\providecommand \bibinfo [0]{\@secondoftwo}%
\providecommand \bibfield [0]{\@secondoftwo}%
\providecommand \translation [1]{[#1]}%
\providecommand \BibitemOpen[0]{}%
\providecommand \bibitemStop [0]{}%
\providecommand \bibitemNoStop [0]{.\EOS\space}%
\providecommand \EOS [0]{\spacefactor3000\relax}%
\providecommand \BibitemShut [1]{\csname bibitem#1\endcsname}%
\bibitem{Nehaniv_07}%
  \BibitemOpen
  \bibfield{author}{%
  \bibinfo {author} {\bibfnamefont{C.~L.}\ \bibnamefont{Nehaniv}}\ and\
  \bibinfo {author} {\bibfnamefont{K.}~\bibnamefont{Dautenhah}},\ }%
  in\ \emph{\bibinfo {booktitle} {Imitation and Social Learning in Robots,
  Humans and Animals}},\ \bibinfo {editor} {edited by\ \bibinfo {editor}
  {\bibfnamefont{C.~L.}\ \bibnamefont{Nehaniv}}\ and\ \bibinfo {editor}
  {\bibfnamefont{K.}~\bibnamefont{Dautenhah}}}\ (\bibinfo {publisher}
  {Cambridge University Press},\ \bibinfo {address} {Cambridge, UK},\ \bibinfo
  {year} {2007})\ pp.\ \bibinfo {pages} {1--18}%
  \bibAnnoteFile{NoStop}{Nehaniv_07}%
\bibitem{Bloom_01}%
  \BibitemOpen
  \bibfield{author}{%
  \bibinfo {author} {\bibfnamefont{H.}~\bibnamefont{Bloom}},\ }%
  \emph{\bibinfo {title} {Global Brain: The Evolution of Mass Mind from the Big
  Bang to the 21st Century}}\ (\bibinfo {publisher} {Wiley},\ \bibinfo
  {address} {New York},\ \bibinfo {year} {2001})%
  \bibAnnoteFile{NoStop}{Bloom_01}%
\bibitem{Moore_96}%
  \BibitemOpen
  \bibfield{author}{%
  \bibinfo {author} {\bibfnamefont{B.~R.}\ \bibnamefont{Moore}},\ }%
  in\ \emph{\bibinfo {booktitle} {Social learning in animals: The roots of
  culture}},\ \bibinfo {editor} {edited by\ \bibinfo {editor}
  {\bibfnamefont{C.~M.}\ \bibnamefont{Heyes}}\ and\ \bibinfo {editor}
  {\bibfnamefont{B.~G.}\ \bibnamefont{Galef}, \bibfnamefont{Jr}}}\ (\bibinfo
  {publisher} {Academic Press},\ \bibinfo {address} {San Diego, CA},\ \bibinfo
  {year} {1996})\ pp.\ \bibinfo {pages} {245--265}%
  \bibAnnoteFile{NoStop}{Moore_96}%
\bibitem{Fiorito_92}%
  \BibitemOpen
  \bibfield{author}{%
  \bibinfo {author} {\bibfnamefont{G.}~\bibnamefont{Fiorito}}\ and\ \bibinfo
  {author} {\bibfnamefont{P.}~\bibnamefont{Scotto}},\ }%
  \bibfield{journal}{%
  \bibinfo {journal} {Science}\ }%
  \textbf{\bibinfo {volume} {256}},\ \bibinfo {pages} {545} (\bibinfo {year}
  {1992})%
  \bibAnnoteFile{NoStop}{Fiorito_92}%
\bibitem{Laland_11}%
  \BibitemOpen
  \bibfield{author}{%
  \bibinfo {author} {\bibfnamefont{K.~N.}\ \bibnamefont{Laland}}, \bibinfo
  {author} {\bibfnamefont{N.}~\bibnamefont{Atton}},\ and\ \bibinfo {author}
  {\bibfnamefont{M.~M.}\ \bibnamefont{Webster}},\ }%
  \bibfield{journal}{%
  \bibinfo {journal} {Phil. Trans. R. Soc. B}\ }%
  \textbf{\bibinfo {volume} {366}},\ \bibinfo {pages} {958} (\bibinfo {year}
  {2011})%
  \bibAnnoteFile{NoStop}{Laland_11}%
\bibitem{Heyes_12}%
  \BibitemOpen
  \bibfield{author}{%
  \bibinfo {author} {\bibfnamefont{C.}~\bibnamefont{Heyes}},\ }%
  \bibfield{journal}{%
  \bibinfo {journal} {Phil. Trans. R. Soc. B}\ }%
  \textbf{\bibinfo {volume} {367}},\ \bibinfo {pages} {2181} (\bibinfo {year}
  {2012})%
  \bibAnnoteFile{NoStop}{Heyes_12}%
\bibitem{Bandura_62}%
  \BibitemOpen
  \bibfield{author}{%
  \bibinfo {author} {\bibfnamefont{A.}~\bibnamefont{Bandura}},\ }%
  in\ \emph{\bibinfo {booktitle} {Nebraska Symposium on Motivation}},\ \bibinfo
  {editor} {edited by\ \bibinfo {editor} {\bibfnamefont{M.~R.}\
  \bibnamefont{Jones}}}\ (\bibinfo {publisher} {University of Nebraska Press},\
  \bibinfo {address} {Lincoln, NE},\ \bibinfo {year} {1962})\ pp.\ \bibinfo
  {pages} {211--274}%
  \bibAnnoteFile{NoStop}{Bandura_62}%
\bibitem{Bandura_77}%
  \BibitemOpen
  \bibfield{author}{%
  \bibinfo {author} {\bibfnamefont{A.}~\bibnamefont{Bandura}},\ }%
  \emph{\bibinfo {title} {Social learning theory}}\ (\bibinfo {publisher}
  {Prentice Hall},\ \bibinfo {address} {New York},\ \bibinfo {year} {1977})%
  \bibAnnoteFile{NoStop}{Bandura_77}%
\bibitem{Kennedy_95}%
  \BibitemOpen
  \bibfield{author}{%
  \bibinfo {author} {\bibfnamefont{J.}~\bibnamefont{Kennedy}}\ and\ \bibinfo
  {author} {\bibfnamefont{R.}~\bibnamefont{Eberhart}},\ }%
  in\ \emph{\bibinfo {booktitle} {Proceedings of the IEEE International
  Conference on Neural Networks}},\ Vol.~\bibinfo {volume} {4}\ (\bibinfo
  {publisher} {IEEE Computer Society},\ \bibinfo {address} {Washington, DC},\
  \bibinfo {year} {1995})\ pp.\ \bibinfo {pages} {1942--1948}%
  \bibAnnoteFile{NoStop}{Kennedy_95}%
\bibitem{Bonabeau_99}%
  \BibitemOpen
  \bibfield{author}{%
  \bibinfo {author} {\bibfnamefont{E.}~\bibnamefont{Bonabeau}}, \bibinfo
  {author} {\bibfnamefont{M.}~\bibnamefont{Dorigo}},\ and\ \bibinfo {author}
  {\bibfnamefont{G.}~\bibnamefont{Theraulaz}},\ }%
  \emph{\bibinfo {title} {Swarm Intelligence: From Natural to Artificial
  Systems}}\ (\bibinfo {publisher} {Oxford University Press},\ \bibinfo
  {address} {Oxford, UK},\ \bibinfo {year} {1999})%
  \bibAnnoteFile{NoStop}{Bonabeau_99}%
\bibitem{Kennedy_98}%
  \BibitemOpen
  \bibfield{author}{%
  \bibinfo {author} {\bibfnamefont{J.}~\bibnamefont{Kennedy}},\ }%
  \bibfield{journal}{%
  \bibinfo {journal} {J. Conflict Res.}\ }%
  \textbf{\bibinfo {volume} {42}},\ \bibinfo {pages} {56} (\bibinfo {year}
  {1998})%
  \bibAnnoteFile{NoStop}{Kennedy_98}%
\bibitem{Fontanari_10}%
  \BibitemOpen
  \bibfield{author}{%
  \bibinfo {author} {\bibfnamefont{J.~F.}\ \bibnamefont{Fontanari}},\ }%
  \bibfield{journal}{%
  \bibinfo {journal} {Phys. Rev. E}\ }%
  \textbf{\bibinfo {volume} {82}},\ \bibinfo {pages} {056118} (\bibinfo {year}
  {2010})%
  \bibAnnoteFile{NoStop}{Fontanari_10}%
\bibitem{Clearwater_91}%
  \BibitemOpen
  \bibfield{author}{%
  \bibinfo {author} {\bibfnamefont{S.~H.}\ \bibnamefont{Clearwater}}, \bibinfo
  {author} {\bibfnamefont{B.~A.}\ \bibnamefont{Huberman}},\ and\ \bibinfo
  {author} {\bibfnamefont{T.}~\bibnamefont{Hogg}},\ }%
  \bibfield{journal}{%
  \bibinfo {journal} {Science}\ }%
  \textbf{\bibinfo {volume} {254}},\ \bibinfo {pages} {1181} (\bibinfo {year}
  {1991})%
  \bibAnnoteFile{NoStop}{Clearwater_91}%
\bibitem{Averbach_80}%
  \BibitemOpen
  \bibfield{author}{%
  \bibinfo {author} {\bibfnamefont{B.}~\bibnamefont{Averbach}}\ and\ \bibinfo
  {author} {\bibfnamefont{O.}~\bibnamefont{Chein}},\ }%
  \emph{\bibinfo {title} {Problem Solving Through Recreational Mathematics}}\
  (\bibinfo {publisher} {Freeman},\ \bibinfo {address} {San Francisco},\
  \bibinfo {year} {1980})%
  \bibAnnoteFile{NoStop}{Averbach_80}%
\bibitem{Hunter_76}%
  \BibitemOpen
  \bibfield{author}{%
  \bibinfo {author} {\bibfnamefont{J.~A.}\ \bibnamefont{Hunter}},\ }%
  \emph{\bibinfo {title} {Mathematical Brain Teasers}}\ (\bibinfo {publisher}
  {Dover},\ \bibinfo {address} {New York},\ \bibinfo {year} {1976})%
  \bibAnnoteFile{NoStop}{Hunter_76}%
\bibitem{Reza_91}%
  \BibitemOpen
  \bibfield{author}{%
  \bibinfo {author} {\bibfnamefont{R.}~\bibnamefont{Abbasian}}\ and\ \bibinfo
  {author} {\bibfnamefont{M.}~\bibnamefont{Mazloom}},\ }%
  in\ \emph{\bibinfo {booktitle} {Proceedings of the Second International
  Conference on Computer and Electrical Engineering}},\ Vol.~\bibinfo {volume}
  {2}\ (\bibinfo {publisher} {IEEE Computer Society},\ \bibinfo {address}
  {Washington, DC},\ \bibinfo {year} {2009})\ pp.\ \bibinfo {pages} {308--312}%
  \bibAnnoteFile{NoStop}{Reza_91}%
\bibitem{Shibanai_01}%
  \BibitemOpen
  \bibfield{author}{%
  \bibinfo {author} {\bibfnamefont{Y.}~\bibnamefont{Shibanai}}, \bibinfo
  {author} {\bibfnamefont{S.}~\bibnamefont{Yasuno}},\ and\ \bibinfo {author}
  {\bibfnamefont{I.}~\bibnamefont{Ishiguro}},\ }%
  \bibfield{journal}{%
  \bibinfo {journal} {J. Conflict Res.}\ }%
  \textbf{\bibinfo {volume} {45}},\ \bibinfo {pages} {80} (\bibinfo {year}
  {2001})%
  \bibAnnoteFile{NoStop}{Shibanai_01}%
\bibitem{Avella_10}%
  \BibitemOpen
  \bibfield{author}{%
  \bibinfo {author} {\bibfnamefont{J.}~\bibnamefont{Gonz\'alez-Avella}},
  \bibinfo {author} {\bibfnamefont{M.}~\bibnamefont{Cosenza}}, \bibinfo
  {author} {\bibfnamefont{V.~M.}\ \bibnamefont{Egu\'{\i}luz}},\ and\ \bibinfo
  {author} {\bibfnamefont{M.~S.}\ \bibnamefont{Miguel}},\ }%
  \bibfield{journal}{%
  \bibinfo {journal} {New J. Phys.}\ }%
  \textbf{\bibinfo {volume} {12}},\ \bibinfo {pages} {013010} (\bibinfo {year}
  {2010})%
  \bibAnnoteFile{NoStop}{Avella_10}%
\bibitem{Peres_11}%
  \BibitemOpen
  \bibfield{author}{%
  \bibinfo {author} {\bibfnamefont{L.~R.}\ \bibnamefont{Peres}}\ and\ \bibinfo
  {author} {\bibfnamefont{J.~F.}\ \bibnamefont{Fontanari}},\ }%
  \bibfield{journal}{%
  \bibinfo {journal} {Europhys. Lett.}\ }%
  \textbf{\bibinfo {volume} {96}},\ \bibinfo {pages} {38004} (\bibinfo {year}
  {2011})%
  \bibAnnoteFile{NoStop}{Peres_11}%
\bibitem{Axelrod_97}%
  \BibitemOpen
  \bibfield{author}{%
  \bibinfo {author} {\bibfnamefont{R.}~\bibnamefont{Axelrod}},\ }%
  \bibfield{journal}{%
  \bibinfo {journal} {J. Conflict Res.}\ }%
  \textbf{\bibinfo {volume} {41}},\ \bibinfo {pages} {203} (\bibinfo {year}
  {1997})%
  \bibAnnoteFile{NoStop}{Axelrod_97}%
\bibitem{Englemore_88}%
  \BibitemOpen
  \bibfield{author}{%
  \bibinfo {author} {\bibfnamefont{R.}~\bibnamefont{Englemore}}\ and\ \bibinfo
  {author} {\bibfnamefont{T.}~\bibnamefont{Morgan}},\ }%
  \emph{\bibinfo {title} {Blackboard Systems}}\ (\bibinfo {publisher}
  {Addison-Wesley},\ \bibinfo {address} {New York},\ \bibinfo {year} {1988})%
  \bibAnnoteFile{NoStop}{Englemore_88}%
\bibitem{Huberman_90}%
  \BibitemOpen
  \bibfield{author}{%
  \bibinfo {author} {\bibfnamefont{B.~A.}\ \bibnamefont{Huberman}},\ }%
  \bibfield{journal}{%
  \bibinfo {journal} {Physica D}\ }%
  \textbf{\bibinfo {volume} {42}},\ \bibinfo {pages} {38} (\bibinfo {year}
  {1990})%
  \bibAnnoteFile{NoStop}{Huberman_90}%
\bibitem{Wilson_75}%
  \BibitemOpen
  \bibfield{author}{%
  \bibinfo {author} {\bibfnamefont{E.}~\bibnamefont{Wilson}},\ }%
  \emph{\bibinfo {title} {Sociobiology}}\ (\bibinfo {publisher} {Harvard
  University Press},\ \bibinfo {address} {Cambridge, MA},\ \bibinfo {year}
  {1975})%
  \bibAnnoteFile{NoStop}{Wilson_75}%
\bibitem{Dunbar_92}%
  \BibitemOpen
  \bibfield{author}{%
  \bibinfo {author} {\bibfnamefont{R.~I.~M.}\ \bibnamefont{Dunbar}},\ }%
  \bibfield{journal}{%
  \bibinfo {journal} {J Human Evol}\ }%
  \textbf{\bibinfo {volume} {22}},\ \bibinfo {pages} {469} (\bibinfo {year}
  {1992})%
  \bibAnnoteFile{NoStop}{Dunbar_92}%
\bibitem{Vespignani_11}%
  \BibitemOpen
  \bibfield{author}{%
  \bibinfo {author} {\bibfnamefont{B.}~\bibnamefont{Gon\c{c}alves}}, \bibinfo
  {author} {\bibfnamefont{N.}~\bibnamefont{Perra}},\ and\ \bibinfo {author}
  {\bibfnamefont{A.}~\bibnamefont{Vespignani}},\ }%
  \bibfield{journal}{%
  \bibinfo {journal} {PLoS ONE}\ }%
  \textbf{\bibinfo {volume} {6}},\ \bibinfo {pages} {e22656} (\bibinfo {year}
  {2011})%
  \bibAnnoteFile{NoStop}{Vespignani_11}%
\bibitem{Laland_98}%
  \BibitemOpen
  \bibfield{author}{%
  \bibinfo {author} {\bibfnamefont{K.~N.}\ \bibnamefont{Laland}}\ and\ \bibinfo
  {author} {\bibfnamefont{K.}~\bibnamefont{Williams}},\ }%
  \bibfield{journal}{%
  \bibinfo {journal} {Behav. Ecol.}\ }%
  \textbf{\bibinfo {volume} {9}},\ \bibinfo {pages} {493} (\bibinfo {year}
  {1998})%
  \bibAnnoteFile{NoStop}{Laland_98}%
\end{thebibliography}%

\end{document}